# Stabilization of betatron tune in Indus-2 storage ring

*Saroj Jena, S. Yadav, R. K. Agrawal , A.D. Ghodke , Pravin Fatnani and T.A. Puntambekar*

Raja Ramanna Centre for Advanced Technology, Indore, India

***Abstract:***

Indus-2 is a synchrotron radiation source which is operational at RRCAT, Indore; India. It is essentially pertinent in any synchrotron radiation facility to store the electron beam without beam loss. During the day to day operation of Indus-2 storage ring difficulty was being faced in accumulating higher beam current. After examining, it was found that the working point was shifting from its desired value during accumulation. For smooth beam accumulation, a fixed desired tune in both horizontal and vertical plane plays a great role in avoiding the beam loss via resonance process. This demanded a betatron tune feedback system to be put in storage ring and after putting ON this feedback, the beam accumulation was smooth. The details of this feedback and its working principle are described in this paper.

Electronic mail: s_jena@rrcat.gov.in

## 1. INTRODUCTION

Indus-2 synchrotron radiation source, located in Indore, India, started providing the beam time to the synchrotron light users since 2008. This facility is a third generation synchrotron radiation storage ring which provides propitious environment to the user communities for carrying out research in various scientific domains. A pre injector, microtron increases the energy of electron generated by thermal emission from a cathode to 20 MeV. After passing through a transfer line, this electron beam is injected to a booster synchrotron that accelerates the beam to 550MeV in less than a second time. Then the electron beam is extracted from booster ring and injected into the storage ring in which beam current is accumulated to its desired level and after that beam energy is ramped to 2.5 GeV [1-2]. The circumference of Indus-2 storage ring is 172.47 meters long and it has 8-fold symmetry. The lattice is a double-bend achromat with two families of quadrupoles in the arc and three families of the matching quadrupoles in the long straight sections. In total there are 72 quadrupoles to adjust the linear optics and 32 sextupoles for correcting chromaticity are distributed over the ring. For each quadrupole family of Q1, Q2, and Q3, there are 16 quadrupole magnets connected with 8 quadrupole power supplies and each family of Q4 & Q5 of 16 quadrupoles are connected with single power supply. The sextupoles are classified by two families and excited by two power supplies; one for each family consists of 16 sextupoles. There are few long straight sections available for the installation of the insertion devices such as wiggler and undulator magnets. Beta functions in both the planes and dispersion function in horizontal plane in one superperiod for the operating lattice of Indus-2 ring is shown in figure 1 and the table 1 shows the general parameters of the ring.

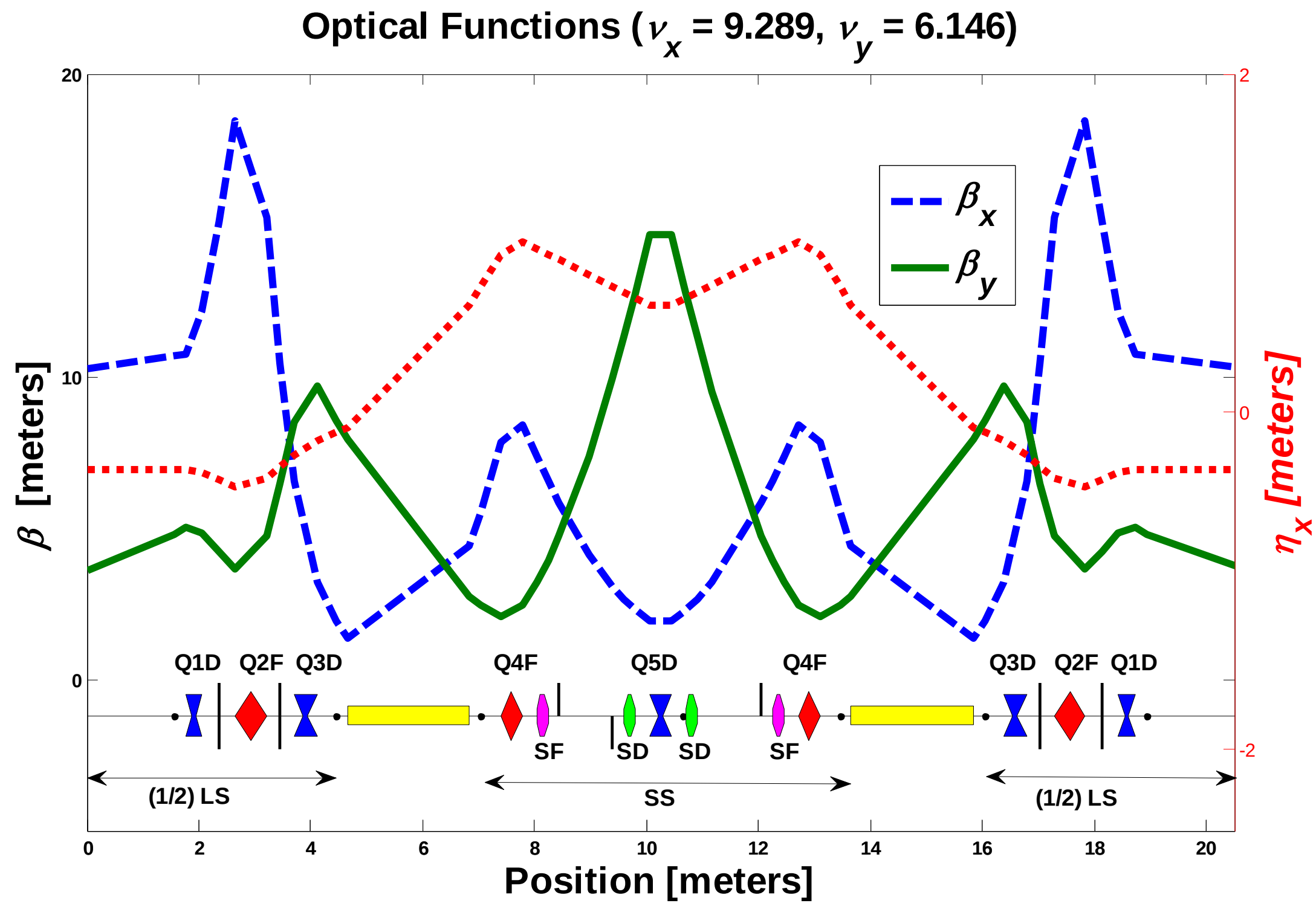

Figure 1: Lattice parameter in one superperiod of Indus-2 storage ring. Q1D, Q3D & Q5D: Defocusing quadrupole magnets. Q2F & Q4F: Focusing quadrupole magnets. SF: Focusing sextupole. SD: defocusing sextupole. LS: Long straight section. SS: Short straight section

Table 1 Beam Parameters of Indus-2

| | |
|---|---|
| Energy | 0.550-2.5 GeV |
| Circumference | 172.47 m |
| Beam current | 300 mA |
| Achromatic structure | DBA |
| Number of unit cell | 8 |
| Betatron tune | 9.29, 6.14 |
| Natural chromaticity | -19, -11 |
| Emittance | 134nm rad at 2.5GeV |
| Coupling coefficient | 0.1 |
| Harmonic Number | 291 |
| RF frequency | 505.8 MHz |
| Critical Photon energy | 6 keV |

## 2. BETATRON TUNE AND RESONANCE

In a storage ring betatron tune [3] is defined as number of betatron oscillations executed by an electron beam in travelling one turn around the ring, which is symbolized by 'ν'.

$$\nu = \frac{1}{2\pi} \oint \frac{ds}{\beta(s)} \tag{1}$$

the integral is taken over the circumference of the ring and β is the betatron amplitude which varies over the length of the ring. The tune value has one integer part & one fractional part and the later is more important as it has a strong impact on beam properties. Betatron tunes depend on the beam optics which is mainly governed by quadrupole focusing strength. The betatron tune gets shift due to a small change Δk in the focusing strength and the shift in tune is given by

$$\Delta\nu = \frac{1}{4\,\pi} \int \beta\,(s)\,\Delta k(s)\,ds \tag{2}$$

In a synchrotron or storage ring, betatron tunes need to remain constant during machine operation which otherwise may drift and cause beam loss via resonance process. Resonance of some kind occurs when the tunes of horizontal and vertical plane, $\nu_x$ & $\nu_z$ satisfy

$$\mathrm{m}\,\nu_x + \mathrm{n}\,\nu_z = \mathrm{P} \tag{3}$$

where m, n and P all are integer. The order of resonance is defined as sum of absolute values of m and n. This order determines the strength of the resonance and impact on the beam which decreases as the order of resonance increases. When P is multiple of machine superperiodicity, the resonance excitation is sensitive to systematic errors in the lattice elements. These resonances are called structural resonances and likely to be strong. The family of resonance lines represented by the above equation fills the whole betatron tune space and it is difficult to find a working point for the machine operation which is sufficiently free from low order resonances. In addition, all resonances have some thickness called stop band width which further reduces the safe space. The stop band width depends on the alignment precision of the magnets and magnetic field errors. Resonances with $\nu_{x,z} = P$ are driven by magnetic field imperfections of dipoles. If any of the horizontal or vertical betatron tunes is an integer value, a dipole magnet imperfection will lead to a transverse kick of the electron orbit each time the electron passes the respective magnet. The kicks add up every turn and therefore lead to increasing oscillation amplitudes until the electron hits the beam pipe. Similarly, quadrupole field imperfections drive resonances with $2\nu_{x,z} = P$ and sextupole fields primarily drive resonances with $3\nu_{x,z} = P$ etc. The resonance lines are the manifestation of imperfections of the magnetic fields and are harmful for the survival of electron beam, if the operating tune point satisfies the resonance equation. When the operating tune point is placed close to a resonance line, the electron beam trajectories in phase space becomes distorted and beam becomes unstable which eventually be lost or scraped by the machine aperture. This phenomenon affects the beam lifetime and injection efficiency.

A small variation of quadrupole strength will cause large betatron oscillations, if the respective betatron tune is close to an integer or half integer resonance. The maximum focusing error, $\Delta k/k$ that can be tolerated may be as small as a few $10^{-3}$ depending on the distance of the betatron tunes to the next resonance line. But the focusing errors in synchrotrons may be much larger due to various physical effects such as imperfections in magnet, hysteresis effect and saturation of magnetic fields etc... Therefore, betatron tunes of ramped storage ring like Indus-2 is usually stabilized by some kind of dynamic magnet field correction which is made by betatron tune feedback mechanism.

## 3. POSSIBLE REASONS OF TUNE SHIFT

Betatron tune plays a crucial role in the performance of a synchrotron light source. In reality, a variety of different sources can result in drift of actual tune value and they are mentioned below.

- Gradient errors in quadrupole magnets
- Closed orbit perturbations and misalignment of sextupole magnet
- Hysteresis in the magnets of machine which employ energy ramping
- Interaction of electrons and residual gas molecules

If a quadrupole magnet has an offset or misaligned w.r.to the electron beam central trajectory, the beam will experience an unwanted kick and generate the closed orbit distortion that can result into tune shift. Horizontal orbit deviation in chromaticity correcting sextupoles creates a shift in tune in both the planes due to the feed down effect; i.e beam experiences a quadrupolar field when it goes off center horizontally at sextupole magnet. Since the beam based alignment is not yet performed in Indus-2, there may be a significant orbit offset present at sextupole magnet. Indus-2 operation involves two energy regimes; i.e electron beam is accumulated at injection energy of 550MeV and then ramped to its maximum designed energy of 2.5GeV. As a result of hysteresis phenomenon, the residual magnetic field effect on betatron tune is noticeable. To cancel out this impact on tune, procedure of cycling of magnets is adopted and after cycling three times, it was observed the effect is minimized. Electron beam circulating through the vacuum chamber ionizes the residual gas molecules present in the chamber. Under certain conditions the positive ions are trapped in the negative potential well of the beam and this phenomenon is called ion trapping. Electron bunch experiences a focusing force from the accumulated trapped ions and this also causes the betatron tune shift.

## 4. REQUIREMENT OF TUNE FEEDBACK

In Indus-2 storage ring betatron tune varies during beam current accumulation as observed from measured data and this tune variation is the reason for poor rate of beam accumulation. The measured tune value at injection energy in Indus2 storage ring for operation of several days is shown in figure 2. It was observed that the tune varies up to 0.01 in a single run and from run to run, this value goes up to 0.02. These variations in tune during beam injection bring hurdles to smooth beam accumulation. It was observed that maintaining the fractional part of tune at [0.278, 0.152] with allowable variation of +/- 0.001 facilitated smooth beam accumulation. This tune point is away from disturbing low order resonance lines and hence doesn't pose a potential threat to efficient operation. As this tune variation is random in nature, feed forward correction doesn't hold good to control the tune. Thus a tune feedback is required to automatically correct the tune values when they go off from the desired values. The main task of a feedback is to control random, non predictable fluctuations in the accelerator parameters, which may be due to internal or external noise. The control of the tunes involves a clean measurement of the tunes and setting up a proper feedback. This helps in stabilizing the machine optics during the machine operation.

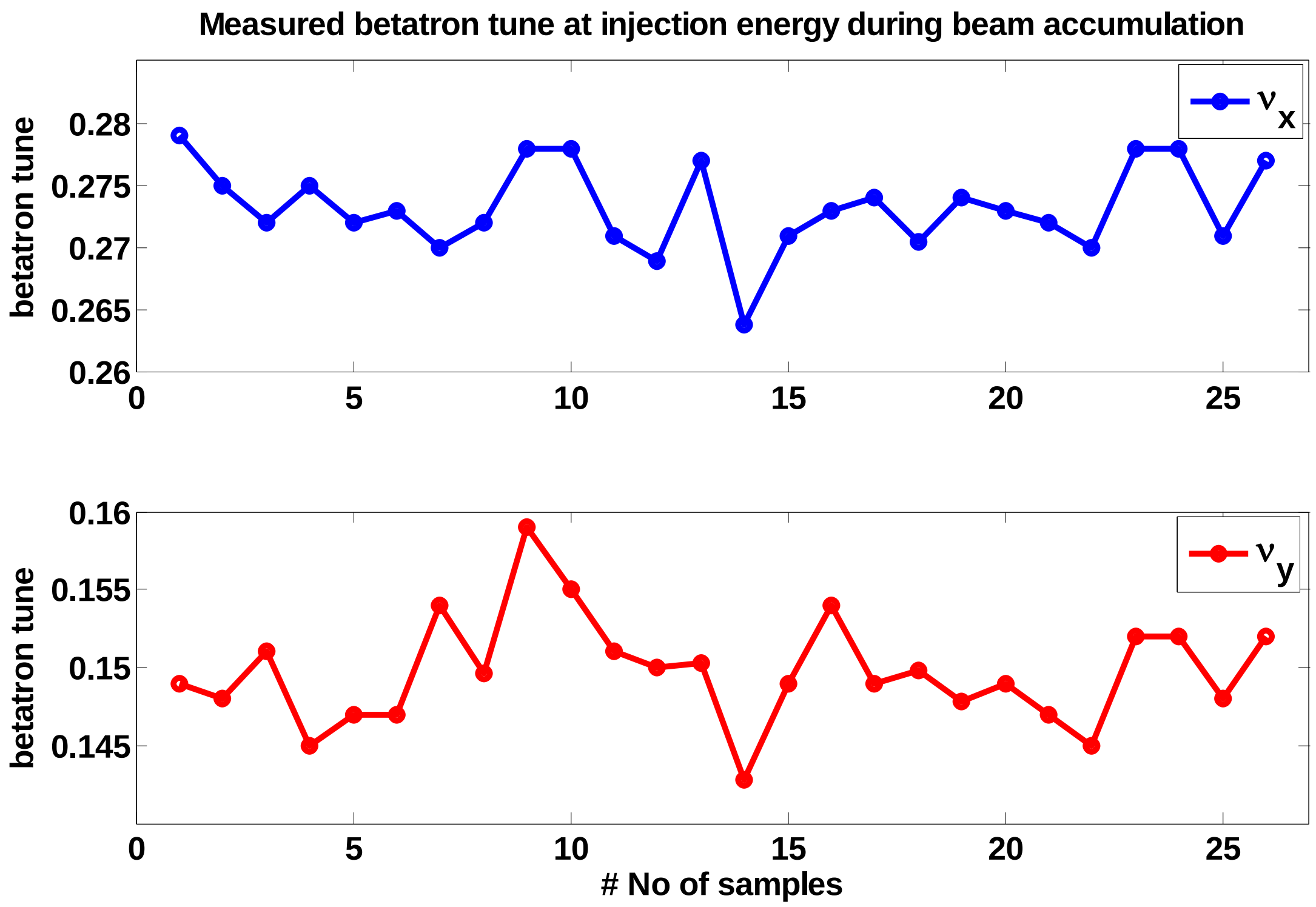

Figure 2: The betatron tune variation at injection energy for several beam accumulations. Upper part: Horizontal tune, Lower part: Vertical tune

## 5. TUNE FEEDBACK SYSTEM FOR INDUS-2

Indus-2 tune measurement system is based on continuous harmonic excitation method [4]. The measurement system employs a spectrum analyzer equipped with a tracking generator. A PC through GPIB bus controls the spectrum analyzer to apply the required settings and acquire the spectrum data. The span and resolution bandwidth of the spectrum analyzer were set to 1.738 MHz and 1 kHz respectively. The resolution of this tune measurement system amounts to 0.0005 tune units. To automate the measurement process, control program with MATLAB based graphical user interface (GUI) has been developed [5]. The schematic diagram of tune feedback control loop [6] is shown in figure 3. The System bandwidth is 0.1 Hz and the residual variation is ±0.0005. The required changes in quadrupole are obtained using the PI control to make system stable and fast converging. The PI coefficients are optimized during tune feedback operation using Ziegler–Nichols method.

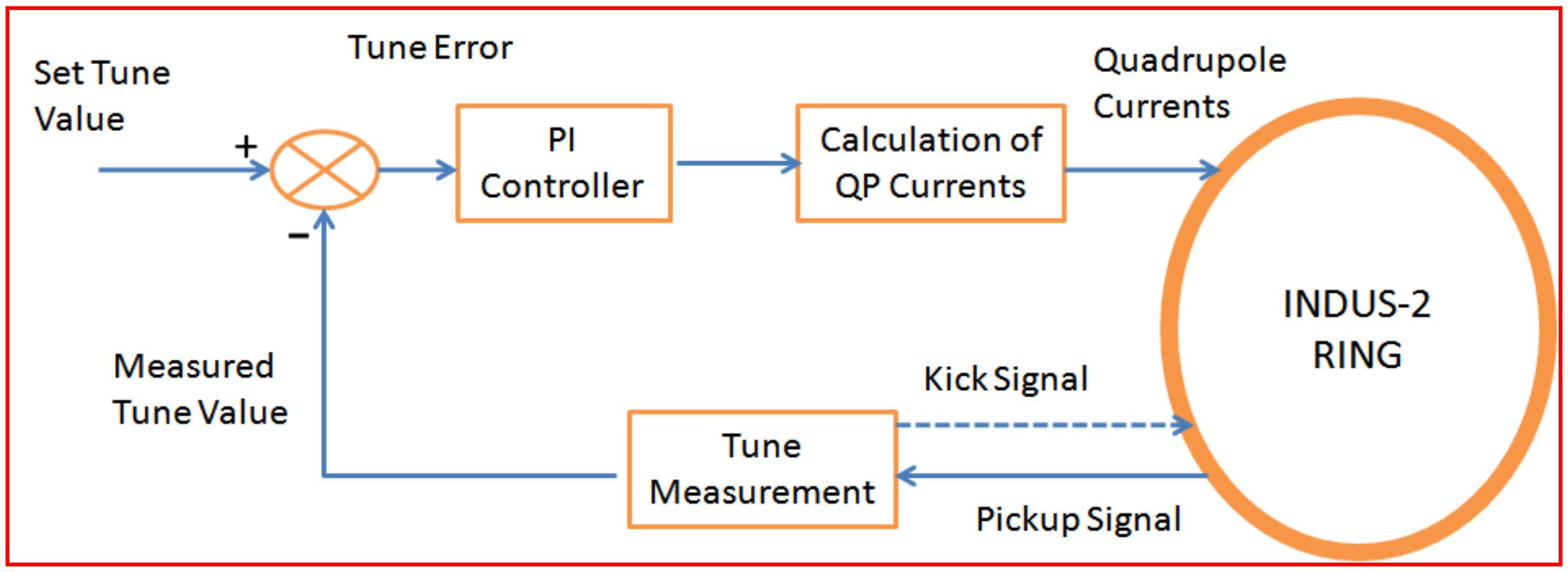

Figure 3: Schematic diagram of Tune feedback system

## 6. CHOICE OF QUADRUPOLE FAMILY FOR BETATRON TUNE CORRECTION

To correct the betatron tune in horizontal and vertical planes, minimum two quadrupole families are required, one is focusing type and another is defocusing type. Here we restrict the maximum number of quadrupole family to be used is two out of five quadrupole families. As Q4F and Q5D families are located in the achromat those quadrupole families are not used for this purpose. The sensitivity of rest of the 3 quadrupole families was checked by measuring betatron tune w.r.to change in current of quadrupole family one by one in Indus-2 and two of them are shown in figure 4. The measured data reveals that per unit change in Q1D current produces very less tune shift in both the plane and thus is not very sensitive to tune. Hence out of the five families two quadrupole families are judiciously decided for use in tune feedback and they are Q2F and Q3D, i.e. these two quadrupoles produce maximum tune change with minimum change in their current. The quadrupole families chosen for this purpose are in the non dispersive arc and are most appropriate because of their decoupled tune sensitivities. Using the real machine, the tune response matrix was calculated by measuring the change in betatron tune in both the plane with reference to change in quadrupole current. In the measurements, the kick size was chosen to create a minimum detectable tune change by the tune measurement system so as to have good noise to signal level. This is measured for several times and the results are consistent with each other.

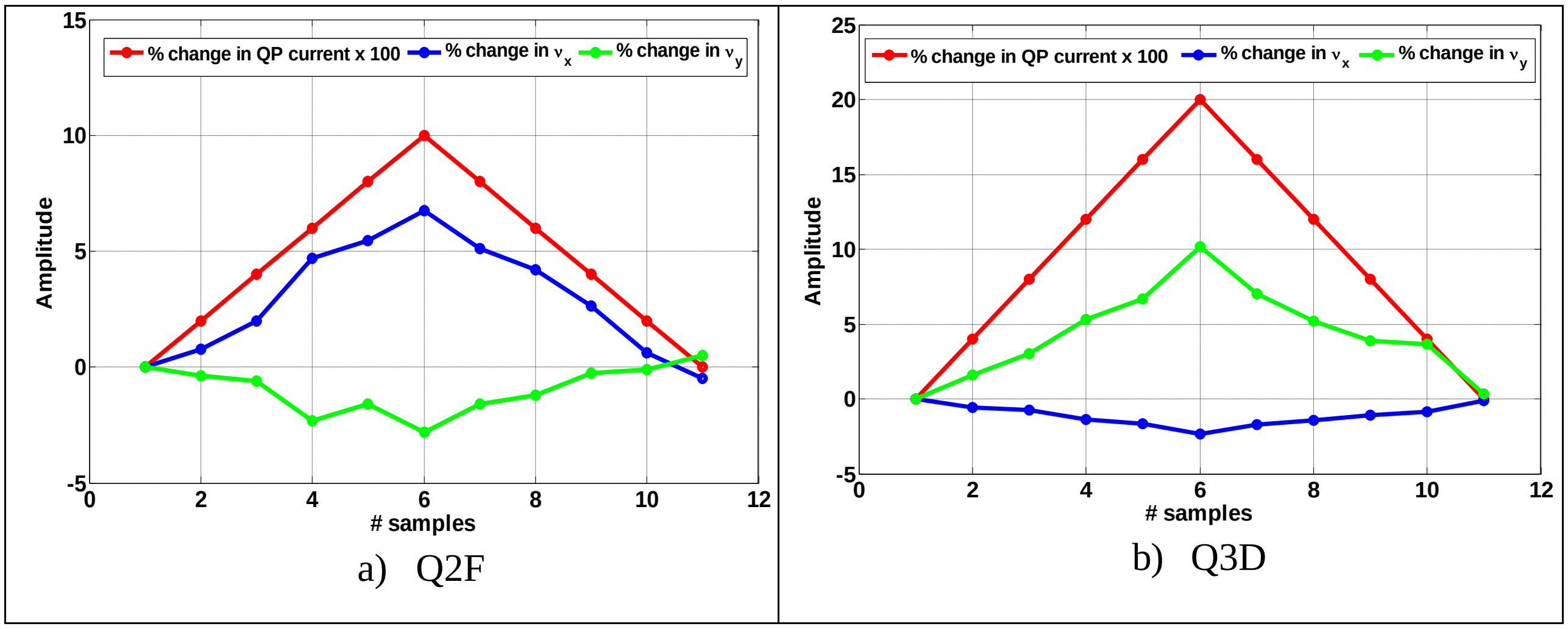

Figure 4: Quadrupole sensitivity to betatron tune for the quadrupole families Q2F & Q3D. Red line: percentage change in Quadrupole current × 100. Blue line: Percentage change in horizontal tune. Green line: Percentage change in vertical tune

## 7. RESULTS AND OBSERVATION

Software code is developed for the tune feedback system which runs on the real operating machine. The code has been written in order to permit the operator to set the betatron tunes to any desired values, using quadrupoles in the non-dispersive straight sections. A graphical user interface (GUI) is made which displays the resonance diagram up to 6$^{th}$ order with the measured tune values flashing on it which facilitates to visualize how far the measured tune point is away from the nearby dangerous resonance lines. The program is written in MATLAB environment which provides an excellent and interactive GUI and also interfaces to hardware of the machine. The main philosophy of the design was based on creating panels which would be user friendly as much as possible with the required safety features. The graphical user interface for the program displays the measured tune and reads the power supplies currents of all the quadrupoles from the machine. An operator just by editing the desired tune value can start the program to perform betatron tune corrections on the machine. The measured tune value ($\nu_x$, $\nu_y$) is subtracted from desired tune value ($\nu_{xREF}$, $\nu_{yREF}$) to obtain tune error ($\Delta\nu_x$, $\Delta\nu_y$). PI control rule is applied to multiply the tune error and the inverse of the pre measured tune transfer matrix to calculate the required changes in the quadrupole power supply currents [7-8]. For small changes of excitation currents of quadrupoles, the variation of tunes depends linearly on the current increments. The linear equation between the change in tune and change in quadrupole current is mentioned below.

$$\overrightarrow{\Delta \nu} = A * \overrightarrow{\Delta I} \qquad (4)$$
$$\overrightarrow{\Delta I} = A^{-1} * \overrightarrow{\Delta \nu}$$

$$\begin{pmatrix} \Delta\nu_x \\ \Delta\nu_y \end{pmatrix} = \begin{pmatrix} a_{11} & a_{12} \\ a_{21} & a_{22} \end{pmatrix} \begin{pmatrix} \Delta I_{QF} \\ \Delta I_{QD} \end{pmatrix} \qquad (5)$$

$$A = \begin{bmatrix} \frac{\Delta\nu_x}{\Delta I_{QF}} & \frac{\Delta\nu_x}{\Delta I_{QD}} \\ \frac{\Delta\nu_y}{\Delta I_{QF}} & \frac{\Delta\nu_y}{\Delta I_{QD}} \end{bmatrix}$$

Using the tune transfer matrix for Indus-2 the coefficient matrix at injection energy of 550 MeV is measured and is given as

$$A = \begin{bmatrix} 0.744694 & -0.18032 \\ -0.1031 & 0.248232 \end{bmatrix} \qquad (6)$$

The tune feedback system calculates the required delta currents to be added or subtracted in Q2F and Q3D quadrupole families. The desired tune is approached in several steps with 20% of full correction in each step. Each step output is displayed in the mesh of resonance lines. The betatron tune is measured once every 10 seconds and correction through feedback in both the planes is applied all of which takes nearly 30 seconds. The tune values during beam current accumulation was plotted in resonance diagram for both the mode, i.e. tune feedback OFF & ON and are shown in figure 5. This figure shows that without the feedback, betatron tune is getting shifted and it crosses the 3rd, 4th and 5th order resonance lines [9]. The equations of these resonance lines are

$$-\nu_x + 2\nu_y = 3 \qquad (7)$$

$$4\nu_x - \nu_y = 31 \qquad (8)$$

$$2\nu_x + 3\nu_y = 37 \qquad (9)$$

$$3\nu_x + \nu_y = 34 \qquad (10)$$

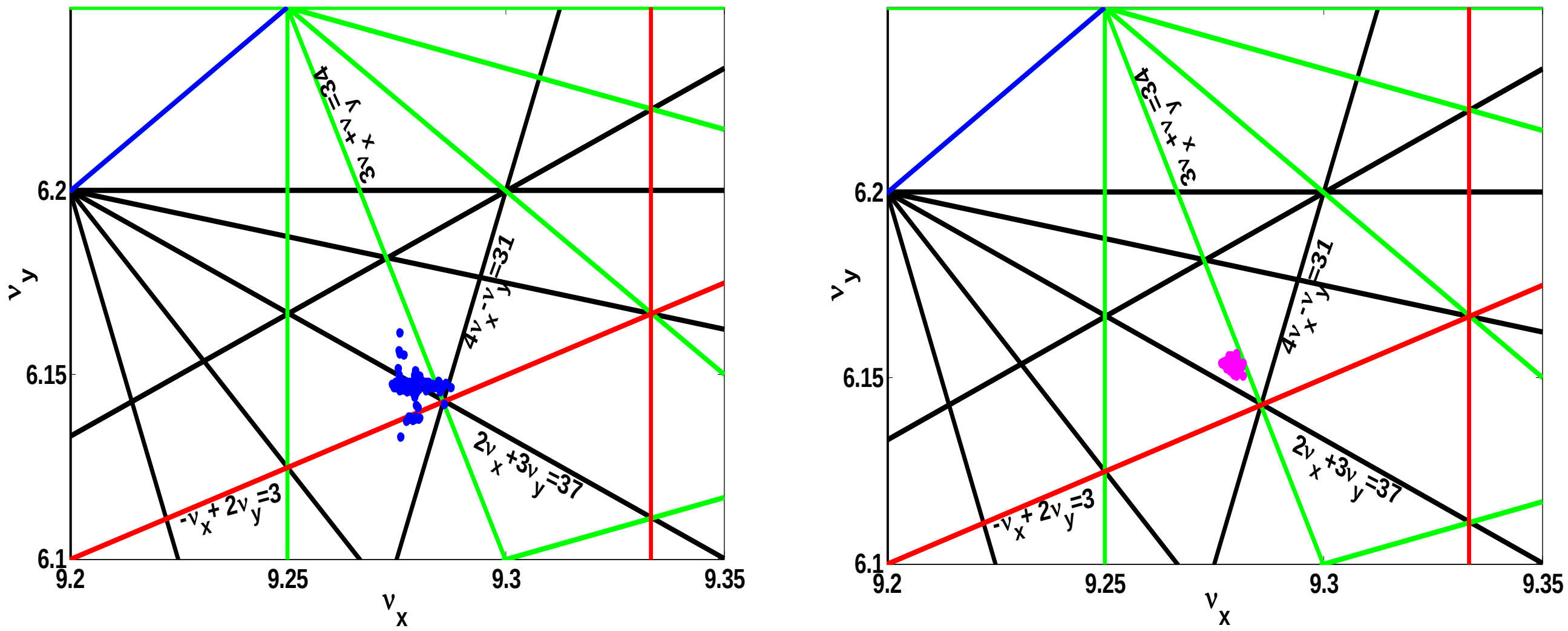

a) Tune feedback OFF    b) Tune feedback ON

Figure 5: Variation of betatron tune during beam current accumulation with tune feedback OFF and ON with resonance diagram upto 5th order. Blue line: 2nd order resonance, Red line: 3rd order resonance, Green line: 4th order resonance, Black line: 5th order resonance

Out of these resonance lines, the 3rd order one written in equation (7) is stronger and partial beam loss occurs while the tune crosses this line shown in red color in figure 5. Near this difference resonance, the coupling transfers energy from horizontal to vertical motion and thus amplitude in vertical plane increases periodically and those particles are eventually lost at physical boundary of vacuum chamber. This 3rd order resonance shouldn't be driven if the 8-fold symmetry of Indus-2 lattice were perfect. However we found this resonance troublesome, which indicated that sizable amount of sextupole errors were present in Indus-2 storage ring. With the feedback put ON, the tune shift is under control and away from all resonance lines shown in figure 5(b) and smooth beam accumulation occurs. From the figure 5(b) it can be seen, the tune point momentarily crosses the fourth order resonance line, at that instant it was found that beam accumulation rate becomes poor but no partial beam loss has occurred. This may be due to the weak strength of resonance line being a higher order one.

Before the installation of the tune feedback systems, betatron tunes was corrected manually by iterative modification of quadrupole current. In case of rapid changes of the betatron tunes, it was not possible to correct the tunes with sufficient precision, and the correction often showed an oscillatory pattern. The tune feedback provides a real-time correction of tune drifts. In Indus-2 storage ring, one typical beam current accumulation at 550MeV is shown in figure 6 and measured tune in both horizontal & vertical plane is shown as the beam current accumulation increased up to 110mA. From the figure it is clearly observed that when tune feedback is OFF, the tune variation is large and creates trouble in beam accumulation. After putting ON the tune feedback, the beam accumulation is smooth and at the same time the tune variation is less in comparison to earlier case. Also with addition of tune feedback system, the beam injection rate has improved by 20% over that without the tune feedback.

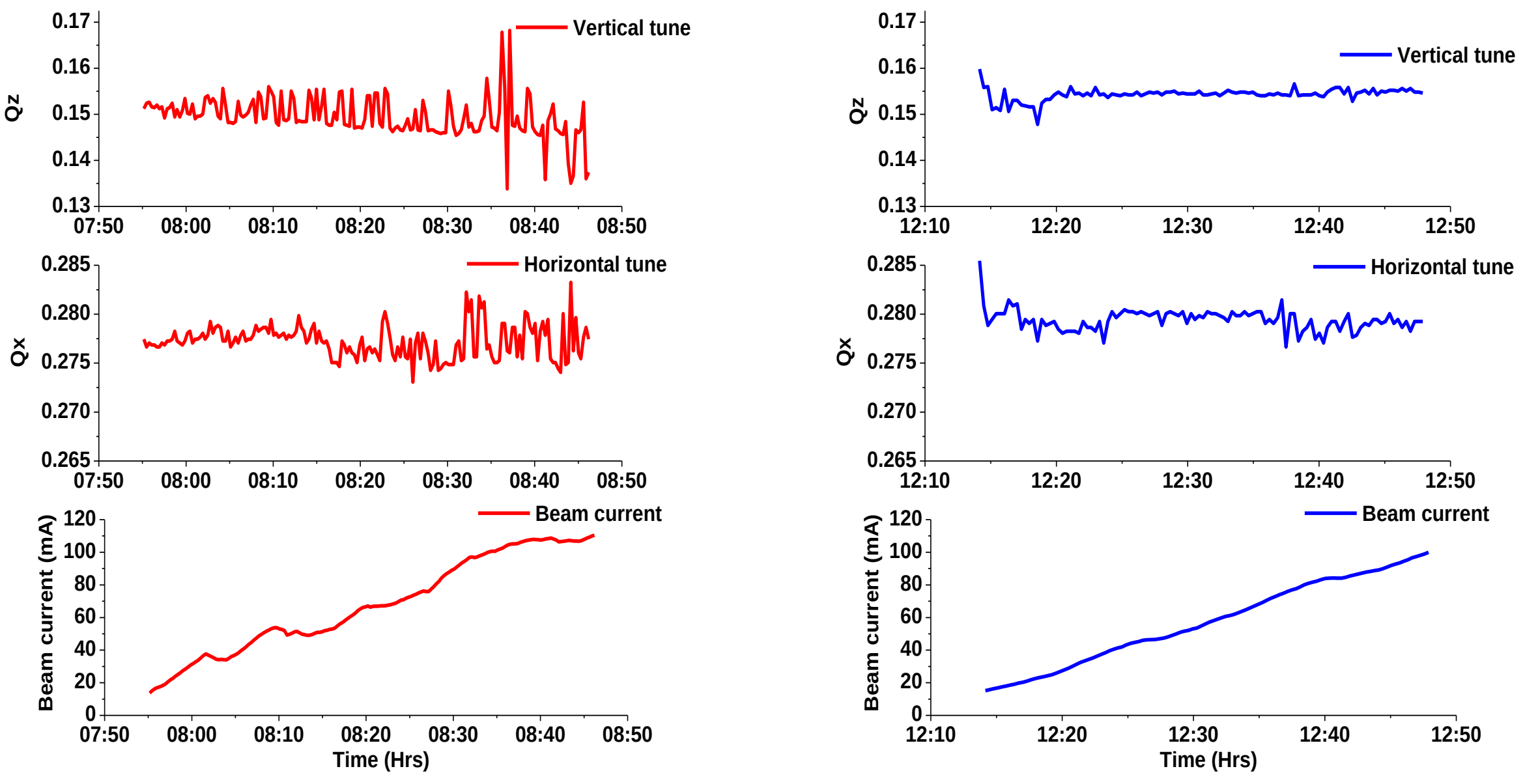

**a)** Tune feedback OFF **b)** Tune feedback ON

Figure 6: Beam current accumulation with tune feedback ON and OFF

At the final energy of 2.5GeV, stability and reproducibility of tune point is utmost important to maintain a beam orbit that provides customary source point location to the Indus-2 beam line users. It was observed the drift in tune at 2.5GeV is small which is of the order of 0.002 and that doesn't create any trouble to users so far. However in future, after the introduction of insertion devices (IDs) in storage ring, the tune feedback will be required at final energy too. The existing tune feedback method will need a modification as the tune value has to be fixed irrespective of the gap opening in several IDs. That time the tune variations will be compensated by the adjustment of the lattice optics, either local compensation of tune that corrects the tune shift by only powering of adjacent quadrupoles or global compensation that uses several quadrupole of the ring. Also in future there is a plan to upgrade the tune measurement system so that the measurement and correction will take place in faster rate than the existing one.

## 8. CONCLUSION

The tune feedback system is developed and tested in Indus-2 and is demonstrated to reduce the tune shift during the beam accumulation at beam injection energy of 550 MeV. The application of tune feedback has proven to be extremely useful during the machine operation, giving an excellent control of the machine and at the same time facilitating operations, which otherwise might have been tedious and time consuming. This tune feedback controls the betatron tune in both the plane within the range of 0.001 and without tune feedback, the maximum tune shift occurs up to 0.01. Presently, the time interval between every correction in the tune feedback is 30 sec which is rather large. However it doesn't make trouble during beam accumulation, but the time between successive corrections needs to be reduced when the feedback will be operational during energy ramping.

***Acknowledgements***

The authors thank P R Hannurkar and Dr. P D Gupta for their fruitful discussion and support to complete this work. They also thank the shift crew members of the Indus accelerator complex for their helps during experiments conducted in Indus-2 storage ring.